\documentclass[twocolumn, amsmath, amssymb, aps]{revtex4-2}
\usepackage[dvipsnames]{xcolor}
\usepackage{graphicx}
\usepackage{enumitem}
\usepackage{siunitx}
\usepackage{booktabs}
\usepackage{multirow}
\usepackage{mathtools}
\usepackage[switch]{lineno}

\newcommand\myshade{30}
\colorlet{mylinkcolor}{red}
\colorlet{mycitecolor}{orange}
\colorlet{myurlcolor}{orange}

\usepackage[%
bookmarks=true,
colorlinks,
linkcolor=mylinkcolor!\myshade!black,
urlcolor=myurlcolor!\myshade!black,
citecolor=mycitecolor!\myshade!black,
plainpages=false,
pdfpagelabels,
final,
breaklinks=true
]{hyperref}

\usepackage[normalem]{ulem}
\newcommand{\clsout}[1]{\textcolor{red!80!black}{\sout{}}}

\usepackage{physics}
\usepackage{braket}

\begin{document}
\title{Realization of three and four-body interactions between momentum states  in a cavity through optical dressing }

\author{Chengyi Luo, Haoqing Zhang, Chitose Maruko, Eliot A.~Bohr, Anjun Chu, Ana Maria Rey, James K.~Thompson}
 \email{jkt@jila.colorado.edu}
 \affiliation{JILA, NIST, and Department of Physics, University of Colorado, Boulder, CO, USA.}
\date{\today}

 \begin{abstract}
Paradigmatic spin Hamiltonians in condensed matter and quantum sensing typically utilize pair-wise or 2-body interactions between constituents in the material or ensemble.  However, there is growing interest in exploring more general $n$-body interactions for $ n >2$,  with examples including more efficient quantum gates or the realization of exotic many-body fracton states.  Here we realize an effective $n=3$-body Hamiltonian interaction using an ensemble of laser-cooled atoms in a high finesse optical cavity with the pseudo-spin 1/2 encoded by two atomic momentum states.  To realize this interaction, we apply two dressing tones that coax the atoms to exchange photons via the cavity to realize a virtual 6-photon process, while the lower-order interactions destructively interfere.  The resulting photon mediated interactions are not only $n>2$-body but also  all-to-all(-to-all)  and therefore of great  interest for fast entanglement generation and quantum simulation of exotic phases such as the long sought but not yet observed charge-Qe superconductors, with $Q=2n$ . The versatility of our experimental system can also allow for extending to 3-body interactions in multi-level systems or to higher-order interactions, such as the signature of a $n=4$-body interaction mediated by a virtual  eight photon process that we also observe.
\end{abstract}

\maketitle

Physical systems can often be described by pair-wise interactions between  objects. However, higher-order interactions, in which  three or more objects  directly couple, appear in many different contexts.
For instance, higher-order  interactions have been theoretically predicted to play a key role in understanding complex biological and social networks \cite{battiston2021physics,grilli2017higher}, in quantum information 
\cite{CombesRotationalError2024,toffoli1980reversible,nielsen2010quantum,zhang2024fast}, molecular and chemical physics \cite{threebodyRMP2013,buchler2007three}, in  condensed matter \cite{BernardPSpin1980,BernardPSpin1981,PoggiDeutsch2023PRXQuantum,Roger2005,Berg2009}, nuclear physics \cite{Fujita1957}, and in high energy physics \cite{Pinto2024}.  Multi-body interactions cannot be decomposed into a sum of pair-wise interactions, but instead emerge from effective theories in which intermediate states can be integrated out to consider only dynamics within a restricted Hilbert space. Experimental challenges include not only implementing the $n$-body interaction in a genuine many-particle system with more than a few atoms, but  also eliminating the lower-order interactions so that the $n$-body interactions are dominant. 

Regardless of the great and broad  interest, there has been limited progress towards engineering controllable $n$-body interactions with $n>2$, with only a few  experimental implementations using for example, superconducting qubits~\cite{fedorov2012implementation, kim2022high,eriksson2024universal}, trapped ions~\cite{katz2022n,katz2023programmable,katz2023demonstration,băzăvan2024squeezing,saner2024generating}, ultra-cold  gases~\cite{dai2017four,Goban2018,Will2010,kraemer2006evidence, Hadzibabic20173bodycontact}, and Rydberg atoms~\cite{levine2019parallel,AhnRydberg2020,kim2024realization}.   Scaling of these multi-body interactions for quantum simulation and sensing to larger systems is important but difficult. Therefore,  their realization in a scalable system such as an optical cavity is highly desirable, but has not yet been achieved.

\begin{figure*}[!thb]
	\centering
	\includegraphics[width=\textwidth]{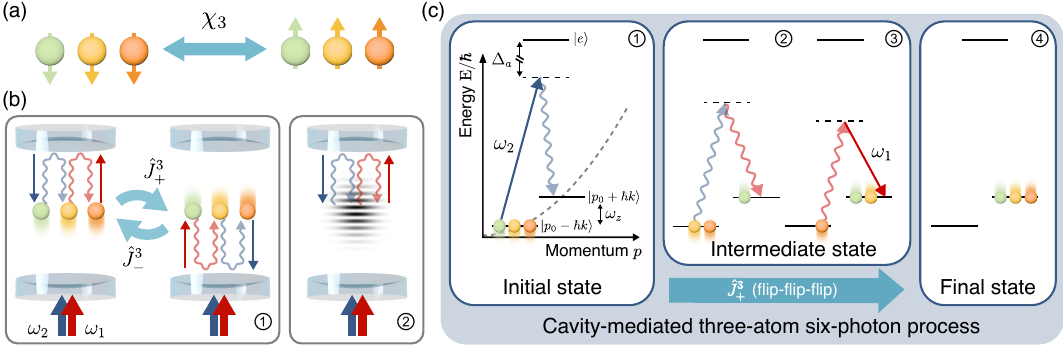}
	\caption{ \textbf{Three-body interaction.}
	    \textbf{(a)} A 3-body lowering and raising interaction causes a trio of spins to flip their spin states simultaneously. \textbf{(b)}  Here, the pseudo-spin is a pair of atomic momentum states $\ket{p_0\pm\hbar k}$ along the cavity, separated by two photon-recoil momenta $2\hbar k$. Left panel: the momentum relative to the average momentum $p_0$ is indicated by the tail on each colored atom. 
        The cavity is driven by two dressing lasers (red and blue arrows). The all-to-all 3-body interaction is realized by the atoms exchanging photons via the cavity, with the intermediate photons of different frequencies shown (red and blue squiggly arrows).
        Right panel: an atomic density grating (grey) is generated by placing the atoms in a superposition of momentum states, inducing multiple back-reflections of light that lead to the 3-body interaction.
        \textbf{(c)} The energy level diagram for the 3-atom and 6-photon process illustrates how three atoms flip their momentum states by showing the intermediate virtual states. Only the relevant fields are shown at each step for simplicity. The kinetic energy difference of the two momentum states is $\hbar \omega_z$, and the pumps are detuned by approximately $\Delta_a$ from $\ket{e}$.  %Since both $\Delta_a\gg \Gamma,\kappa, \omega_z$ and the intermediate photons (squiggly) are non-resonant with the cavity mode,
        We operate in detuned regimes such that both the intermediate optical excited state and the intermediate photon states (squiggly arrows) can be adiabatically eliminated to arrive at the 3-body pseudo-spin interaction. 
        }
	\label{fig1}
\end{figure*}

Here, we experimentally realize a 3-body interaction between a thousand rubidium atoms, each prepared in a superposition of two momentum states that act as pseudo-spin states. The interaction is all-to-all(-to-all) and consists of triplets of atoms all flipping their momentum or pseudo-spin states at once as shown in Fig.~\ref{fig1}(a).  This is achieved by applying  two dressing laser tones along the axis of an optical cavity to realize a resonant 3-atom and 6-photon process with four of the atom-photon interactions mediated by the cavity mode. By setting the power ratio between the dressing laser tones to one, a pure resonant 3-body interaction can be achieved with lower order 2-body interactions (exchange and pair raising/lowering processes) cancelled via symmetry~\cite{ThompsonMomentumExchange2024,luo2024hamiltonian}. The collective nature of the cavity mediated  interactions  is essential to make the thee-body terms  comparable to previously demonstrated 2-body interactions,  since the slower rate imposed by the exchange of six photons is compensated by the fact that the strength of the 3-body term scales  with atom number as $N^2$,  compared to  the  $N^1$ scaling of the 2-body counterpart.

We also demonstrate the extension of our approach to $n$-body with $n>3$. As an example, we observe signatures of a resonant 4-body interaction. Furthermore, in the future with more dressing laser tones applied to the cavity, we can  couple not only two but   $m>2$ atomic momentum or internal states, enabling us to generate collective  $n>2$-body interactions between  $m>2$-level systems, a promising system for the implementation of quantum simulation of complex models including quantum field theories \cite{banuls2020simulating,aidelsburger2022cold}, robust quantum  hybrid quantum processors \cite{liu2024hybridoscillatorqubitquantumprocessors},  and error correcting codes  \cite{Cai2021}.

\vspace{2mm}
\noindent\textbf{Experimental setup.} In the experiment, $^{87}$Rb atoms are laser-cooled inside a vertically-oriented standing wave cavity, see Fig.~\ref{fig1}(b) and \cite{Thompson2022SAI, ThompsonMomentumExchange2024}. The atoms are allowed to fall along the cavity axis, guided by a repulsive radial potential which is provided by a blue-detuned and hollow optical dipole trap. To prepare and manipulate the atomic momentum states, two-photon velocity-selective Raman or Bragg transitions can be driven by non-resonantly injecting pairs of laser beams with different frequencies into the cavity. After accelerating under gravity for 20~ms, we select around $N=1000$ atoms with momentum along the cavity axis $p_0-\hbar k$ and rms momentum spread $\sigma_p<0.1\hbar k$ where $\hbar$ is the reduced Planck constant, the wavenumber is $k=2\pi/\lambda$, and the wavelength is $\lambda = 780$~nm. The atomic internal states are then transferred to the ground hyperfine state $\ket{F=2,m_F=2}$ using sequential microwave pulses.

We then apply the Bragg lasers to prepare the atoms in a superposition of two wave packets with momenta centered on $p_0 \pm \hbar k$, which we use to define a qubit between states $\ket{\uparrow}\equiv \ket{p_0 + \hbar k}$ and $\ket{\downarrow}\equiv \ket{p_0 - \hbar k}$. Ignoring the finite momentum spread, we can then define $\hat{\psi}^\dagger_{\uparrow,\downarrow}$ and $\hat{\psi}_{\uparrow,\downarrow}$ as the operators for creating and annihilating an atom in the momentum states $\ket{\uparrow}$ and $\ket{\downarrow}$. The collective ladder operators $\hat{J}_+ = \hat{\psi}_\uparrow^\dagger\hat{\psi}_\downarrow$ and $\hat{J}_- = \hat{\psi}_\downarrow^\dagger\hat{\psi}_\uparrow$ can then be defined for mapping to a collective pseudo-spin model. Similarly, the collective spin projection operators can be defined as $\hat{J}_x = \frac{\hat{J}_+ + \hat{J}_-}{2}$, $\hat{J}_y = \frac{\hat{J}_+ - \hat{J}_-}{2 i}$ and $\hat{J}_z = \frac{\hat{\psi}^\dagger_\uparrow \hat{\psi}_\uparrow-\hat{\psi}^\dagger_\downarrow\hat{\psi}_\downarrow}{2}$. For all collective spin operators, the expectation values are denoted by $J_\alpha=\langle \hat{J}_\alpha \rangle$ with $\alpha\in [x,y,z,+,-]$.

The two Bragg tones can be described by complex amplitudes in the lab frame $\alpha_{B1}$ and $\alpha_{B2}$ such that the differential phase between the Bragg tones accrues as $\psi_B= \arg(\alpha_{B1}^* \alpha_{B2})= \omega_z t +\phi_B + \frac{1}{2}\delta_r t^2$.  Here, $\omega_z= 2 k v_0$ is the two-photon Doppler shift for photons moving along the cavity axis at speed $v_0= p_0/m_a$ where $m_a$ is the mass of the atom.  The frequency splitting $\omega_z/2 \pi$ is typically about 500~kHz for our experiments, but since the atoms are accelerating due to gravity, we linearly ramp the frequency difference of the two tones at a rate $\delta_r= 2 \pi\times 25.11~$kHz/ms to compensate.  The differential phase $\phi_B$ sets the azimuthal phase of the Bragg rotation on the Bloch sphere.

\begin{figure}[thb!]
	\centering
	\includegraphics[width=0.48\textwidth]{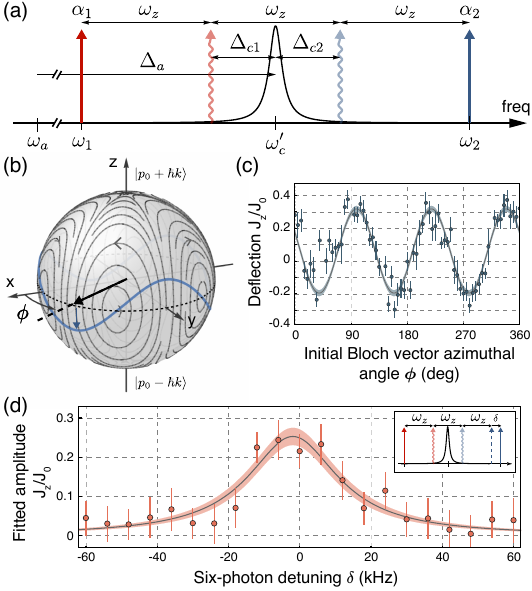}
	\caption{\textbf{Dynamics and spectroscopic signal for three-body interaction.}
  	\textbf{(a)} Frequency diagram. The cavity mode is detuned by $\Delta_a$ from an atomic cycling transition. Two dressing lasers (red and blue arrows) are applied at frequencies $\omega_{2,1}$, with frequency difference $3\omega_z$, and amplitudes $\alpha_{2,1}$. Scattering from the atoms generates the two new sideband tones (red and blue squiggly lines). The four tones combine to resonantly drive the 3-atom and 6-photon process illustrated in Fig.~\ref{fig1}. 
    \textbf{(b)} Numerically simulated mean-field dynamics induced by the 3-body interaction. The flow lines indicate the direction that a Bloch vector prepared at a given point on the Bloch sphere  will subsequently move with time. The blue curve is the predicted short time displacement of the spin projection $J_z$ for Bloch vectors initially prepared at different azimuthal angles on the equator ($\theta=\pi/2$, black dashed line.)
    \textbf{(c)} Experimentally measured displacements of $J_z$ due to the 3-body interaction, normalized by $J_0 = N/2$. The data  displays the predicted three-cycle oscillation as the initial azimuthal angle $\phi$ is varied from 0 to $360\deg$. 
    \textbf{(d)}   
    The three-cycle oscillation of (c) is only observed when the 6-photon process is resonant.  The fitted amplitude of the three-cycle oscillation is shown as a function of the 6-photon detuning $\delta=\omega_1 - \omega_2 - 3\omega_z$, see frequency diagram in inset. Data are fit with 68\% confidence bands and all error bars reported are $1\sigma$ uncertainties.}
	\label{fig2}
\end{figure}

\vspace{2mm}
\noindent\textbf{Three-body interactions with two-tone dressing.} 
As shown in Fig.~\ref{fig2}(a), a cavity mode is stabilized to the blue of the D2 cycling transition $\ket{F=2,m_F=2} \rightarrow \ket{F'=3,m_{F'}=3}$ with a detuning $\Delta_{a}=\omega_c-\omega_a= 2\pi \times 500~\mathrm{MHz}$. The detuning $\Delta_a$ is large compared to all other relevant frequencies including the excited state decay rate $\Gamma=2 \pi \times 6$~MHz and the cavity power decay rate $\kappa = 2\pi \times 56(3)~$kHz. In this far-detuned limit, an atom at position $Z$ shifts the cavity resonance by $\frac{g_0^2}{\Delta_a} \cos^2 (k Z)$, where $ 2 g_0=2\pi \times 0.96~\mathrm{MHz}$ is the maximal single photon Rabi coupling at a cavity anti-node \cite{ZilongsPRA}.

We create the 3-body interaction by injecting two dressing laser tones at frequencies $\omega_1$ and $\omega_2$ into the cavity, as shown by the solid red and blue arrows in Fig.~\ref{fig1}(b,c).  The two tones are separated in frequency by $\omega_2-\omega_1=3\omega_z$, and their amplitudes are described by complex amplitudes $\alpha_1$ and $\alpha_2$, as shown in Fig.~\ref{fig2}(a).

The mechanism by which the 3-body interaction arises is illustrated in Fig.~\ref{fig1}(b,c). A blue dressing laser photon at frequency $\omega_2$ is absorbed and reemited into the cavity (lighter blue squiggly arrow) in the opposite direction with a frequency shift $-\omega_z$ as shown in Fig.~\ref{fig1}(b1,c1) and Fig.~\ref{fig2}(a).  This virtual two-photon process transfers $2\hbar k$ of momentum to the absorbing atom, flipping its pseudo-spin.  This process then repeats twice more but with the frequency-shifted photons being absorbed and back-scattered by other atoms, at each step receiving a frequency shift $-\omega_z$. This virtual process terminates with stimulated emission into the red dressing laser at frequency $\omega_1= \omega_2 - 3\omega_z$, and three different atoms having flipped their momentum or pseudo-spin state. While illustrated with three atoms, we note that each step is collectively enhanced by constructive interference between paths in which different atoms in the $N$ atom ensemble undergo the momentum flip, as compared to for instance, a single atom decreasing its momentum state three times in a row. This constructive interference causes the collective interaction characteristic energy scale to increase as $N^2$.

After adiabatic elimination of the intermediate virtual states in the above, this cavity-mediated process manifests as a collective 3-body raising and lowering processes described by the effective Hamiltonian
\begin{equation}
    \hat{H} = \chi_3 \hat{J}_+^3 e^{3 i \phi_d}+ \chi_3 \hat{J}_-^3 e^{-3 i \phi_d}. 
    \label{eq:H3body}
\end{equation}

\noindent The phase of the interaction $3\phi_d$ is given by  the differential phase of the two dressing tones with respect to the Bragg pulse that creates the initial superposition of momentum states $3\phi_d = \arg \left(\alpha_1^* \alpha_2 \right) - 3\phi_B$ with $\phi_B$ the phase of the Bragg coupling (see SM for details). The interaction strength is given by
\begin{equation}
    \chi_3 =\left(\frac{g_0^2}{4\Delta_a}\right)^3 \left|\alpha_1 \alpha_2\right|
    \mathrm{Re} \Big[ \frac{1}{\left(\Delta_{c1}+ i \kappa/2\right)\left(\Delta_{c2}+ i \kappa/2\right)}\Big] ,
\end{equation}

\noindent where the detuning of the red and blue sideband tones (intermediate squiggly arrows) from resonance with the cavity appear in the denominator as $\Delta_{c2}= \left(\omega_{2}-\omega_z\right) -\omega'_c $ and $\Delta_{c1}= \left(\omega_{1}+\omega_z\right) -\omega'_c$.  Here $\omega'_c= \omega_c + N g_0^2/(2\Delta_a)$ is the time-averaged cavity resonance frequency. The incident dressing laser power on the cavity is typically only a few hundred pW.

Each dressing tone individually also induces a 2-body exchange interaction of the form $\chi_2\hat{J}_+\hat{J_-}$ \cite{luo2024hamiltonian}.  For typical operating conditions we expect $\left|\chi_3/\chi_2\right|\ll 1$.  However, the \emph{collective} interaction strengths are in fact comparable  $\left| \frac{\chi_3 N^2}{\chi_2 N} \right| \approx 1/2$.  Moreover, the 2-body exchange interactions of the two dressing lasers can be tuned to have opposite signs and cancel by simply operating in the symmetric configuration shown in Fig.~\ref{fig2}(a) with  $\Delta_{c2} = -\Delta_{c1}$ and $\left|\alpha_2\right|=\left|\alpha_1\right|$.  The symmetry thus allows the 3-body interactions here to dominate the  observed dynamics.

There is a complementary way to understand the 3-atom and 6-photon process in terms of the atomic spatial density grating shown in Fig.~\ref{fig1}(b2).  The atomic density grating is the modulated quantum probability density of finding the atoms at location $Z$ along the cavity axis due to having prepared the atoms in a superposition of momentum states. The density grating has a periodicity of $\lambda/2$ and appears to the light as a spatially varying index of refraction that can reflect the light, much like a dielectric stack mirror. With the atomic density grating of Fig.~\ref{fig1}(b2) moving along the cavity at velocity $v_0$, a downwards-going blue dressing laser photon (blue solid arrow) is reflected by the atomic density grating.  The reflected light experiences a Doppler shift of precisely $-\omega_z$, which creates the sideband photon (lighter blue squiggly arrow), and at the same time an atom receives a two-photon momentum kick $2\hbar k$, flipping its pseudo-spin state. The sideband photon is then  reflected and frequency shifted by the atomic density grating twice more, flipping the momentum or pseudo-spin state of two other atoms, before the photon is stimulated back into the  red dressing laser (red solid arrow.)

\vspace{2mm}
\noindent\textbf{Mean-field dynamics.} Assuming $\phi_d=0$, the mean-field equations of motion for the collective spin operators can be derived from the effective 3-body interaction Hamiltonian along with Lindblad operators that capture collective decay through the cavity (see SM for details):
\begin{equation}
\begin{aligned}
\dot{J}_{+} & =-6 i \chi_3 J_z J_-^2 - \Gamma J_+ \\
\dot{J}_{z}  & =-3i\chi_3 ( J_+^3- J_-^3) - 2\Gamma J_z.
\end{aligned}
\label{eq:eom}
\end{equation}

Here, $\Gamma$ is the collective or superradiant decay rate, with the ratio to the interaction strength given by $\Gamma/\chi_3 \sim \frac{4\Delta_a \kappa}{g_0^2}$.  In the symmetric configuration used in the experiment, the superradiance is balanced, resulting in no collective enhancement for this process (see SM for details). However, the unitary dynamics driven by the 3-body interaction is enhanced by a factor of $N^2$  (see below) so that $\Gamma/N^2\chi_3 \ll 1 $, and to good approximation we can set $\Gamma=0$ in the subsequent discussion of collective mean-field dynamics.
For extending the discussion to $\phi_d\neq 0$, one can simply apply the substitution $J_\pm \rightarrow J_\pm e^{\pm i \phi_d},J_z\rightarrow J_z$ to Eq.~\eqref{eq:eom}.

Fig.~\ref{fig2}(b) shows the simulated mean-field dynamics induced by the 3-body interactions. The displayed flow lines (grey lines) indicate the trajectory that a Bloch vector follows on the sphere with time~\cite{PoggiDeutsch2023PRXQuantum, borish2020transverse,luo2024hamiltonian}.  
Specifically, the local flow vector is defined by $(\mathbf{J}_f - \mathbf{J}_i)/\Delta t$ for any initial Bloch vector $\mathbf{J}_i/(N/2)=(\sin\theta \cos\phi,\sin\theta\sin\phi,\cos\theta)$ prepared on the Bloch sphere, and with $\mathbf{J}_f$ the final Bloch vector after interacting for a short time $\Delta t$.
%In the experiment, we initialize the system with a vector $\mathbf{J}_i/(N/2)=(\sin\theta \cos\phi,\sin\theta\sin\phi,\cos\theta)$  on the Bloch sphere, evolve it over a short time $\Delta t$, and measure the final state $\mathbf{J}_f$. The local flow vector is then calculated as $(\mathbf{J}_f - \mathbf{J}_i)/\Delta t$.
The resulting flow lines predict eight fixed points on the Bloch sphere: two unstable fixed points located at the north and south poles with $\mathbf{J}_i/(N/2)=(0,0,\pm 1)$ (i.e.~$\theta = 0, \pi$) and six stable fixed points located on the equator (i.e.~$\theta = \pi/2$) and evenly spaced in their azimuthal angles $\phi$ by $\pi/3$. These are consistent with the  Hamiltonian being   symmetric    under a $\pi/6$ rotation around $\hat{z}$ and a time-reversal operation.

In Fig.~\ref{fig2}(b) we consider an ensemble of initial states prepared on the equator ($\theta=\pi/2$, black dashed line) with different azimuthal angles chosen in the range $\phi \in (0,2\pi)$. The blue curve shows the predicted change in spin projection $\Delta \mathbf{J}_{z}/\left(N/2\right)= \frac{3}{2}\chi_3 N^2 \Delta t \sin(3\phi)$ along $\hat{z}$ after the short-time evolution, which reflects the $N^2$ enhancement factor. When the initial Bloch vector is prepared at a stable fixed point, there is no change in the Bloch vector orientation $\Delta \mathbf{J}=0$, here occurring at azimuthal angles $\phi=q \pi/3$ with $q$ an integer.

\begin{figure}[h!]
	\centering
	\includegraphics[width=0.48\textwidth]{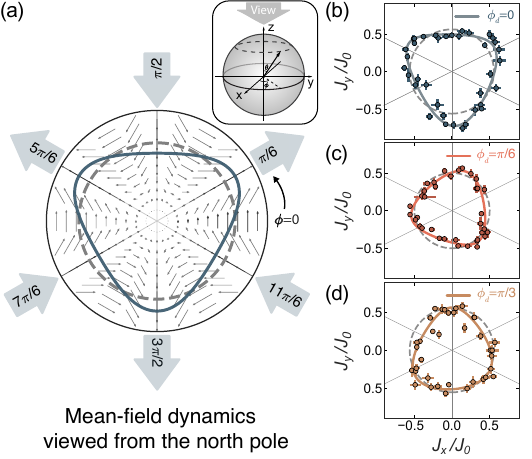}
	\caption{\textbf{Deformation of circular distribution.}
\textbf{(a)} Theoretical mean-field flow lines shown as a vector field (light grey arrows) and as viewed from above the north pole (see inset.) The thick arrows surrounding the Bloch sphere highlight the fact that near the trifurcation points the flow lines alternate every $\pi$/3 in azimuthal angle between flowing towards versus away from the unstable trifurcation point at the north pole. As a consequence, after a short time of evolution, a distribution of Bloch vectors prepared at different $\phi$ but constant polar angle $\theta =\pi/4$ illustrated by the dashed grey circle, will evolve for short interaction times into a distribution shown by the light blue curve.
\textbf{(b)} Experimental observation of the predicted mean-field dynamics in \textbf{(a)}. The distribution of final measured Bloch vectors on the $J_x$-$J_y$ plane deforms away from a circular distribution after the 3-body interaction is applied for a short time.
%In the experiment, the distribution of the Bloch vector on the $J_x$-$J_y$ plane is distorted into triangular after the interaction. 
\textbf{(c)-(d)} The final Bloch vector distribution is rotated by $\pi/6$ and $\pi/3$ by repeating the experiment but changing the relative phase $3\phi_d$ between the two dressing tones by $\pi/2$ and $\pi$ respectively.  This shows that the orientation of the 3-body interaction can be  controlled via the dressing laser phases.}
	\label{fig3}
\end{figure}

To understand the unstable fixed points, we project the theoretical flow vector field onto the $\hat{x}-\hat{y}$ plane (grey arrows in Fig.~\ref{fig3}(a)) as viewed looking down at the north pole of the Bloch sphere. Several key directions are highlighted below. From Eq.~\eqref{eq:eom}, we find $\dot{J}_z=0$ at $\phi=q\pi/3$, meaning that the flow lines exhibit only an azimuthal component at these points. Furthermore, the equation for $\dot{J}_+$ can be reformulated in terms of $(\theta,\phi)$, where $\dot{\phi}=-\frac{3}{4}\chi_3 N^2\cos(3\phi) \sin(2\theta)$ implying that the azimuthal angle does not change at initial Bloch vector angles $\phi=\pi/6+q\pi/3$, and only the polar angle changes at these azimuthal angles. Additionally, the expression for $\dot{\theta}=-\frac{3}{2}\chi_3 N^2 \sin^2\theta \sin(3\phi)$ shows that $\dot{\theta}>0$ for $\phi=\pi/6+2q\pi/3$ and $\dot{\theta}<0$ for $\phi=\pi/2+2q\pi/3$. These features define the trifurcation points at the north and south poles, e.g.~near the north pole trifurcation point the flow lines alternate every $\pi/3$ in azimuthal angle between flowing toward ($\dot{\theta}<0$) versus away from ($\dot{\theta}>0$) the trifurcation point, as shown by the thick arrows in  Fig.~\ref{fig3}(a).

\vspace{2mm}
\noindent\textbf{Observing dynamics near the stable fixed points}
To experimentally observe the predicted dynamics, we apply a Bragg $\pi/2$-pulse lasting $30~\mu$s to prepare the initial Bloch vector on the equator (i.e.~$J_z=0$). After $50~\mu$s of separation, a Bragg $\pi$-pulse is applied to refocus the wave packets. We applied the dressing lasers for $50~\mu$s after the wave packets re-overlap with $\chi_3 N^2 = 2\pi\times 390~$Hz. The population in each momentum state is measured at the end to determine the spin projection $J_z$.  We then repeat the measurement sequence for different initial Bloch vector azimuthal angles $\phi$, achieved by changing the Bragg pulse phase $\phi_B$ while holding the dressing laser phase $3\phi_d$ fixed.  Fig.~\ref{fig2}(c) shows that the measured spin projection $J_z$ oscillates up and down three times as $\phi_B$ and hence $\phi$ is varied 0 to $2 \pi$.  The six azimuthal angles at which maximum deflections occur are points halfway between adjacent fixed points where the quasi-circular flows move in the same directions.  The six points between maximum and minimum deflection occur at azimuthal angles that are aligned to the stable fixed points.

We also experimentally confirm that the measured deflections of the spin projection $J_z$ arise from the resonant 6-photon interaction outlined in Fig.~\ref{fig1}.  To do this, we move the two dressing lasers away from the 6-photon resonance condition by setting $\omega_2-\omega_1 = 3\omega_z + \delta$.  The frequency $\omega_z$ is independently determined via Bragg spectroscopy. (For simplicity in this discussion, we neglect the frequency chirping at rate $\delta_r$ that is still applied as previously described.) We then scan the Bloch vector's azimuthal angle 0 to $2\pi$ and extract the 3-cycle fringe height at a given detuning $\delta$.  As shown in Fig.~\ref{fig2}(d), one sees a resonance in the fringe height near $\delta=0$ as expected and with a FWHM of 28~kHz, determined by the duration of the interaction. 

\vspace{2mm}
\noindent\textbf{Observing dynamics near an unstable fixed point} 
We experimentally measure the  dynamics near the unstable trifurcation point at the north pole by observing the evolution of Bloch vectors prepared at initial azimuthal angles $\phi= 0$ to $2 \pi$ but at a constant polar angle  $\theta=\pi/4$.  %Specifically, we start with all atoms in one momentum state, apply a Bragg $\pi/4$-pulse with the phase $\phi_B$ scanned from 0 to $2\pi$ with a step of $\pi/18$. 
To prespare these initial states, we start with the Bloch vector at the south pole.  We then apply an initial Bragg $\pi/4$-pulse.  After 50~$\mu$s, we apply a Bragg $\pi$-pulse that refocusses the matter-wave packets while also setting the polar angle to $\theta=\pi/4$. The interaction is then applied when the two wave packets are maximally re-overlapped for 50~$\mu$s. A final Bragg $\pi/2$-pulse is applied along $\hat{y}$ or $\hat{x}$ to map the spin projections  $J_x$ or $J_y$  onto $J_z$, the spin projection which we can measure (see SM for details.) 

Without the interaction applied, we expect a circular distribution of final Bloch vectors centered around $\hat{z}$. After evolution under the 3-body interaction, the measured circular distributions is shown in Fig.~\ref{fig3}(b) is seen to deform into a more triangular distribution, consistent with the predicted trifurication point at the north pole where there are three converging flow lines equally interspersed between three diverging flow lines.  We also demonstrate phase control of the interaction by repeating the experiment but changing the phase of the interaction $\phi_d\rightarrow \phi_d+ \pi/6$ and $\phi_d+ \pi/3$.  In Fig.~\ref{fig3}(c) and (d), one sees that the deformation of the originally circular  distribution is also rotated by $\pi/6$ and $\pi/3$ respectively.

\begin{figure}[th!]
	\centering
	\includegraphics[width=0.45\textwidth]{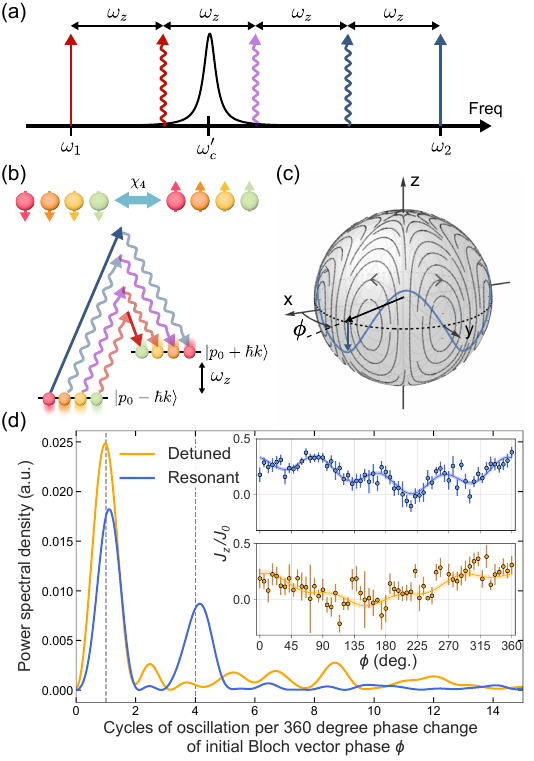}
	\caption{\textbf{Signature of four-body interaction.} 
    \textbf{(a)} Frequency diagram for driving 4-body interaction. The two dressing laser tones (red and blue arrows) are separated by $4\omega_z$ and the intermediate photons (squiggly lines) are equally separated by $\omega_z$. 
    \textbf{(b)} Energy level diagram for the 4-atom and 8-photon process that flips the momentum state of four items simultaneously.
    \textbf{(c)} Theoretical flowlines for the mean-field dynamics. With four stable points  on the equator, the Bloch vector will experience a four-cycle oscillation in the change of the spin projection $J_z$ when scanning the azimuthal phase of the initial Bloch vector.
    \textbf{(d)} (insets) The measured deflection $J_z$ due to applying the 4-body interaction versus the azimuthal phase of the initial Bloch vector. Blue data is with the dressing lasers on resonance for the 8-photon process. Yellow data is for the dressing lasers tuned away from resonance.  One sees a 4 cycle oscillation of $J_z$ in the resonant case. (main) The power spectrum of the data shows a 4-cycle peak at resonance, and no peak when off-resonance.  In contrast, the one-cycle peak does not depend strongly on being resonant or not.
	}  
	\label{fig4}
\end{figure}

\vspace{2mm}
\noindent\textbf{Signature of a 4-body interaction.}
We also realize a 4-body interaction of the form $\hat{H}_4 = \chi_4 \hat{J}_+^4 e^{4 i \phi}+ \chi_4^* \hat{J}_-^4 e^{-4 i \phi}$, and observe a signature of the expected dynamics.  To achieve this, we apply the dressing lasers with a frequency difference $\omega_2-\omega_1=4\omega_z$ (as shown in Fig.~\ref{fig4}(a)) which brings a 4-atom and 8-photon process (Fig.~\ref{fig4}(b)) into resonance.  %\clsout{The resulting Hamiltonian after eliminating the cavity mode  is now $\hat{H}_4 = \chi_4 \hat{J}_+^4 e^{4 i \phi}+ \chi_4^* \hat{J}_-^4 e^{-4 i \phi}$, with $\chi_4 =\left(\frac{g_0^2}{4\Delta_a}\right)^4 
%\frac{\alpha_-}{\Delta_{c_-}-\omega_z + i \kappa/2} \frac{\alpha_+^*}{\Delta_{c_+}+\omega_z + i \kappa/2} \frac{1}{\Delta_{c_+}+2 \omega_z + i \kappa/2}.$} 
The 2-body exchange interactions can again be cancelled by adjusting the relative power in the dressing laser, and the 3-body interaction is non-resonant in this case. Note that the characteristic frequency scale of this interaction is now collectively enhanced as $N^3$, but the larger detuning from cavity resonance of the additional intermediate photon leads to a smaller predicted interaction strength $\chi_4 N^3\approx 2\pi \times 130~$Hz. The predicted mean-field dynamics is illustrated by the calculated flowlines shown in Fig.~\ref{fig4}(c).

To observe dynamics induced by this 4-body interaction, we again prepare the initial Bloch vector on the equator with $\theta=\pi/2$ and measure the change in $J_z$ versus initial Bloch vector azimuthal phase $\phi$. The measured fringe is shown in Fig.~\ref{fig4}(d inset) with the two dressing lasers tuned to resonance (blue) and detuned from resonance (red). One can see a clear 4-cycle oscillation in the displacement of $J_z$ as expected from the 8 stable points on the equator that result from a 4-body interaction.% \clsout{that we believe arises from a spurious rf tone that induces an undesired Bragg coupling}.  

The main figure shows the Fourier power spectra of the insets with a clear peak at the 4-cycle frequency only occurring above the noise floor when the 4-photon coupling is on resonance. %This distinction also rules out the possibility of the detuned Rabi drive coming from the two dressing laser tones, which should not demonstrate resonant behaviour. 
Although not included here, the Fourier spectrum of the oscillation induced by the 3-body interaction in Fig.~\ref{fig2}(c) produces a peak centered at the 3-cycle frequency.

There is also a 1-cycle Fourier component that is also apparent in the inset traces of $J_z$ versus $\phi$. The similar amplitude of the 1 cycle component in the 8-photon resonant versus non-resonant data indicates that this is not related to a resonant interaction between the dressing tones, and $n$-body interactions that depend on a single dressing tone should not depend on the phase of the dressing light. We currently attribute this spectral feature to single particle physics due to an unwanted Bragg coupling driven by a spurious sideband that appears in the generation of the two dressing laser tone.  

\vspace{2mm}
\noindent\textbf{Summary.} 
Here, we demonstrate all-to-all 3-body  and 4-body interactions in a matter-wave-cavity coupled system. 
Our  approach relies on resonant multi-photon and multi-particle processes. By tuning the desired process on resonance and detuning the unwanted processes, pure $n$-body interactions can be generated with the lower-order processes cancelled via symmetry. The $n$-body interaction can be understood as $n$ different atoms   flipping   their momentum states in concert from $p_0-\hbar k$ to $p_0+\hbar k$ and vice versa. The  $n$-body interactions are experimentally observed by both a spectroscopic signal and via  mean-field dynamics measured over the Bloch sphere. 
 Our platform can allow for more energy levels and more atoms, allowing for full connectivity, which would also be interesting to combine with local control in the future~\cite{periwal2021programmable,cooper2024graph,Lev2024PRX}.  While demonstrated using momentum states in a cavity, the interaction can also be applied using internal atomic or molecular energy levels.  

To our knowledge, this is the first $n>2$ interaction generated via photon mediated interactions  in  an optical cavity, which provides a whole new sets of tool for quantum simulation and entanglement generation using higher-order correlations.  This opens the path to a broad host of future explorations of exotic many-body physics from new regimes of self-organization physics \cite{baumann2010dicke, Kroezelev2018} to quantum simulation of charge-4e  superconductors \cite{Berg2009}, as well as quantum metrology with new classes of entangled states.  It will also be of great interest to understand in future work how single-particle and collective dissipation~\cite{ThompsonMomentumExchange2024,wilson2024entangled} (predominantly described by jump operators $\hat{J}_-$) compete with entanglement arising from the $n$-body Hamiltonian dynamics. Furthermore, even though we have discussed the implementation of these schemes in optical cavities, our protocols can be straight-forwardly generalized to other types of systems where interactions are mediated by a bosonic mode such as trapped ions or superconducting qubits. Since different platforms are affected by different types of decoherence processes,  it will be interesting to understand  optimal  performance in the different settings.

\begin{acknowledgments}
We acknowledge helpful feedback on the manuscript from Nelson Darkwah Oppong and Allison Carter.
This material is based upon work supported by the U.S. Department of Energy, Office of Science, National Quantum Information Science Research Centers, Quantum Systems Accelerator. We acknowledge additional funding support from the National Science Foundation under Grant Numbers 1734006 (Physics Frontier Center) and  OMA-2016244 (QLCI Q-SEnSE), the Vannevar-Bush Faculty Fellowship, the Heising-Simons foundation and NIST.  

\end{acknowledgments}

% \bibliography{Three_Body}
%
\clearpage
\newpage

\onecolumngrid
\begin{center}
    \textbf{\Large{Supplemental Material}}
\end{center}

\makeatother
\renewcommand{\theequation}{S\arabic{equation}}
\setcounter{equation}{0}
\renewcommand{\thesection}{S\arabic{section}}
\renewcommand{\thetable}{S\arabic{table}}
\setcounter{table}{0}
\renewcommand{\thefigure}{S\arabic{figure}}
\setcounter{figure}{0}
\renewcommand{\bibnumfmt}[1]{[S#1]}   
\renewcommand{\citenumfont}[1]{S#1}
\setcounter{enumiv}{0}   

% \begin{bibunit}[unsrt]  
% \putbib[Three_Body_Supplement] 
% \end{bibunit}

\section{Derivation of the  effective three-body Hamiltonian}
\subsection{Second-order perturbation theory}

\begin{figure*}[th!]
\centering
\includegraphics[width=0.55\textwidth]{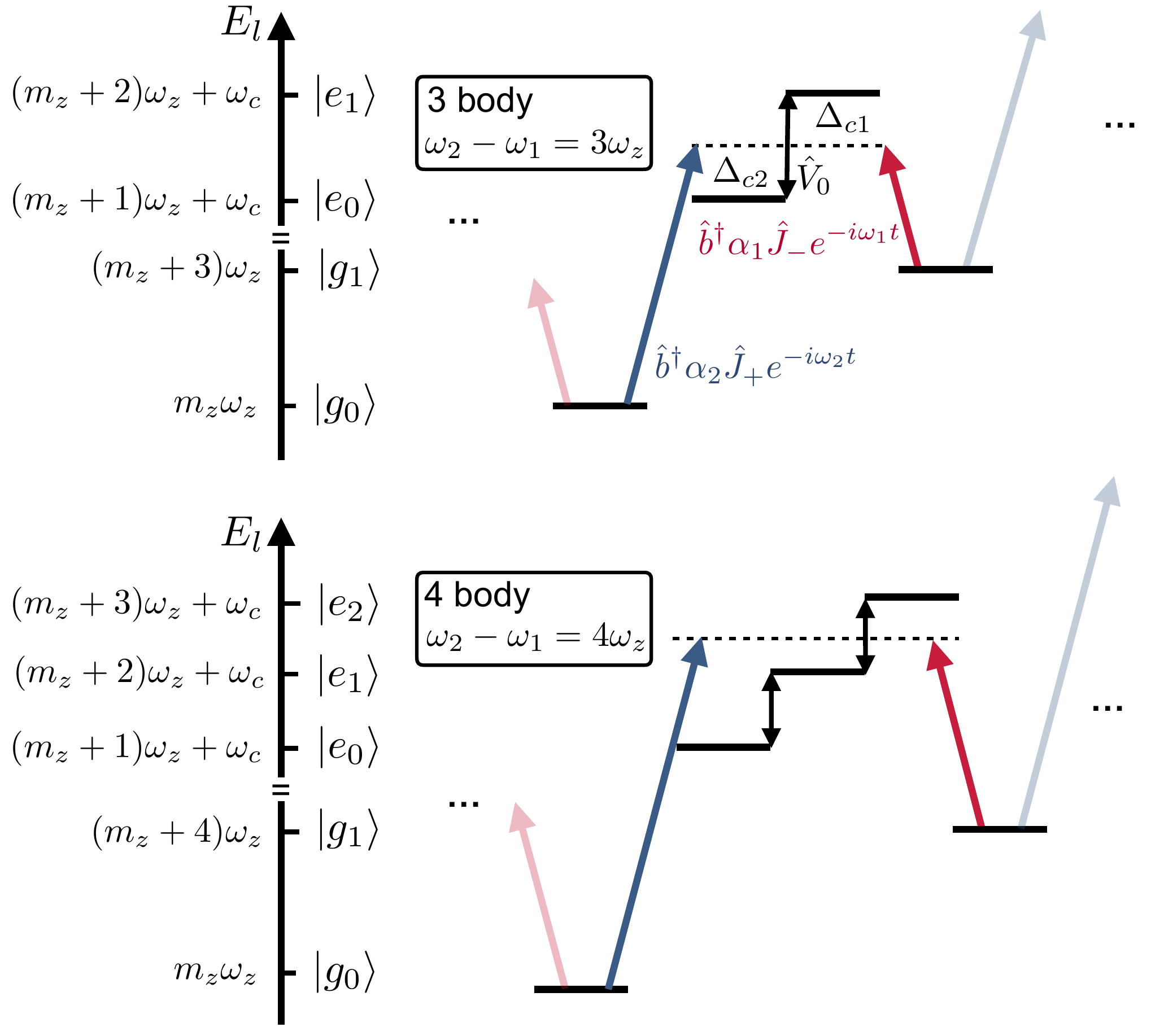}
\caption{\textbf{Diagrammatic perturbation theory for multi-body interactions.} Upper panel: 3-body interaction mediated by a 6-photon process involving two intermediate states. The effective coupling process starts with the excitation of state $\ket{g_0}$ to $\ket{e_0}$ through the absorption of a blue pump photon. This is followed by the coupling between $\ket{e_0}$ and $\ket{e_1}$ via the operator $\hat{V}_0$, and finally, the system transitions to $\ket{g_1}$ through the emission of a red pump photon. Lower panel: 4-body interaction mediated by an 8-photon process involving three intermediate states.
	}  
	\label{fig5}
\end{figure*}

 In this section, we derive the 3-body interaction Hamiltonian presented in Eq.~\eqref{eq:H3body}. After adiabatic elimination of the excited state~\cite{ThompsonMomentumExchange2024,luo2024hamiltonian,ReyZhang2023BO}, the system Hamiltonian is given by:
\begin{equation}
\begin{aligned}
\hat{H}_S &= \hat{H}_0 + \frac{g_{0}^{2}}{4\Delta_{a}}\hat{a}^{\dagger}\hat{a}\left(\hat{J}_{+}+\hat{J}_{-}\right) \\
    &+ \left(\epsilon_{1}e^{-i\omega_{1}t} + \epsilon_{2}e^{-i\omega_{2}t} \right)\hat{a}^{\dagger} + \rm h.c., \label{eq:hlab}
\end{aligned}
\end{equation}
here $\hat{H}_{0}=\omega_{c}\hat{a}^{\dagger}\hat{a}+\omega_{z}\hat{J}_{z}$. The  applied two dressing lasers have  amplitudes $\epsilon_{1,2}$ at frequencies  $\omega_{1,2}$. We define $\mathcal{G}=\frac{g_0^{2}}{4\Delta_{a}}$ for simplicity and substitute the cavity field operator via:
\begin{equation}
\hat{a}=\hat{b}+\alpha_{1}e^{-i\omega_1 t}+\alpha_{2}e^{-i\omega_2 t}
\label{eq:sub}
\end{equation}
with the oscillating classical field built in the cavity  given by $\alpha_{1,2}=\frac{\epsilon_{1,2}}{i\kappa/2+(\omega_{1,2}-\omega_c)}$ and $\hat{b}$ representing the cavity field quantum fluctuation.
The atom-cavity coupling Hamiltonian is then given by $\hat{V}=\hat{V}_{+}+\hat{V}_{-}+\hat{V}_{0}$, where
\begin{equation}
\begin{aligned}
\hat{V}_{+}&=\mathcal{G}\hat{b}^{\dagger}\left(\alpha_{1}e^{-i\omega_{1}t}+\alpha_{2}e^{-i\omega_{2}t}\right)\left(\hat{J}_{+}+\hat{J}_{-}\right)\\
\hat{V}_{-}&=\left(\hat{V}_{+}\right)^{\dagger}\quad\hat{V}_{0}=\mathcal{G}\hat{b}^{\dagger}\hat{b}\left(\hat{J}_{+}+\hat{J}_{-}\right)
\end{aligned}
\end{equation}
as the perturbation fields.
Additionally, the cavity dissipation is described by $\hat{L}=\sqrt{\kappa} \hat{b}$. 

We work with the basis states $\ket{m_{z},n_{b}}$, where $n_{b}$ represents the photon number in the cavity mode (with $\hat{b}^{\dagger}\hat{b}\ket{m_{z},n_{b}} = n_{b}\ket{m_{z},n_{b}}$). Here $m_{z}$ is the eigenvalue of the collective pseudo-spin operator  along the $z$-axis (with $\hat{J}_{z}\ket{m_{z},n_{b}} = m_{z}\ket{m_{z},n_{b}}$). Initially, we start in the state $\ket{m_{z},n_{b}=0}$, corresponding to the ground state manifold with zero photons in mode $\hat{b}$. The operator $\hat{V}_+$ couples this state to the intermediate state manifold where $n_{b}=1$. We apply the general formula for second-order perturbation theory, as derived in Ref.~\cite{ReitherSorensen2012effective}, which involves multiple perturbation fields and non-perturbative coupling within the ground state manifold (with $\omega_{z} \sim \left|\omega_{c} - \omega_{1,2}\right|$), resulting in the following effective Hamiltonian and jump operator:
\begin{equation}
\begin{aligned}
\hat{H}_{{\rm eff}} & =-\frac{1}{2}\left[\hat{V}_{-}\sum_{f,l}\left(\hat{H}_{{\rm NH}}^{(f,l)}\right)^{-1}\hat{V}_{+}^{\left(f,l\right)}\left(t\right)+\mathrm{h.c.} \right]+\hat{H}_{g}\\
\hat{L}_{{\rm eff}} & =\hat{L}\sum_{f,l}\left(\hat{H}_{{\rm NH}}^{(f,l)}\right)^{-1}\hat{V}_{+}^{\left(f,l\right)}\left(t\right).~\label{eq5}
\end{aligned}
\end{equation}
Here, the index $f$ represents the perturbation fields with frequency
$\omega_{f}$, $l$ denotes the ground state with energy $E_{l}$,
and $\hat{H}_{{\rm NH}}^{\left(f,l\right)}=\hat{H}_{{\rm NH}}-E_{l}-\omega_{f}$ with $\hat{H}_{\rm NH}=\hat{H}_e-\frac{1}{2}\hat{L}^\dagger \hat{L}$ is  the non-Hermitian Hamiltonian for the intermediate state manifold.

We consider a four-level model where $\ket{g_{0}}\equiv\ket{m_{z},0}$, $\ket{e_{0}}\equiv\ket{m_{z}+1,1}$, $\ket{e_{1}}\equiv\ket{m_{z}+2,1}$ and $\ket{g_{1}}\equiv\ket{m_{z}+3,0}$
with the ground states $\ket{g_{0,1}}$ and intermediate state $\ket{e_{0,1}}$ shown in Fig.~\ref{fig5}, therefore the non-Hermition Hamiltonian is given by:
\begin{equation}
\hat{H}_{{\rm NH}}=\left[\left(m_{z}+1\right)\omega_{z}+\omega_{c}-i\frac{\kappa}{2}\right]\ket{e_{0}} \bra{e_{0}} 
+\left[\left(m_{z}+2\right)\omega_{z}+\omega_{c}-i\frac{\kappa}{2}\right]\ket{e_{1}} \bra{e_{1}} 
+\mathcal{G}f_{m_{z}+1}\left(\ket{e_{1}} \bra{e_{0}}+\ket{e_{0}} \bra{e_{1}}\right)
\end{equation}
where $f_{m_{z}}=\sqrt{\frac{N}{2}(\frac{N}{2}+1)-m_{z}(m_{z}+1)}$ representing the coupling factor between Dicke states $\left|m_{z}\right\rangle $
and $\left|m_{z}+1\right\rangle $. Also, we have the perturbation terms:
\begin{equation}
\hat{V}_{+}=\mathcal{G}\left(\alpha_{1}e^{-i\omega_{1}t}+\alpha_{2}e^{-i\omega_{2}t}\right)\left(f_{m_z}\ket{e_{0}} \left\langle g_{0}\right|+f_{m_z+2}\ket{e_{1}} \left\langle g_{1}\right|\right).
\end{equation}

As a result, there exist four $\hat{H}_{\rm NH}^{(f,l)}$ for various frequencies $\omega_f$ and ground levels $l$:
\begin{equation}
\begin{aligned}
\hat{H}_{{\rm NH}}^{(f=1,l=g_{0})} & =\left(-\Delta_{c1}+2\omega_{z}-i\frac{\kappa}{2}\right)\left|e_{0}\right\rangle \left\langle e_{0}\right|+\left(-\Delta_{c1}+3\omega_{z}-i\frac{\kappa}{2}\right)\left|e_{1}\right\rangle \left\langle e_{1}\right|+\mathcal{G}f_{m_{z}+1}\left(\left|e_{1}\right\rangle \left\langle e_{0}\right|+\left|e_{0}\right\rangle \left\langle e_{1}\right|\right)\\
\hat{H}_{{\rm NH}}^{(f=2,l=g_{0})} & =\left(-\Delta_{c2}-i\frac{\kappa}{2}\right)\left|e_{0}\right\rangle \left\langle e_{0}\right|+\left(-\Delta_{c2}+\omega_{z}-i\frac{\kappa}{2}\right)\left|e_{1}\right\rangle \left\langle e_{1}\right|+\mathcal{G}f_{m_{z}+1}\left(\left|e_{1}\right\rangle \left\langle e_{0}\right|+\left|e_{0}\right\rangle \left\langle e_{1}\right|\right)\\
\hat{H}_{{\rm NH}}^{(f=1,l=g_{1})} & =\left(-\Delta_{c1}-\omega_{z}-i\frac{\kappa}{2}\right)\left|e_{0}\right\rangle \left\langle e_{0}\right|+\left(-\Delta_{c1}-i\frac{\kappa}{2}\right)\left|e_{1}\right\rangle \left\langle e_{1}\right|+\mathcal{G}f_{m_{z}+1}\left(\left|e_{1}\right\rangle \left\langle e_{0}\right|+\left|e_{0}\right\rangle \left\langle e_{1}\right|\right)\\
\hat{H}_{{\rm NH}}^{(f=2,l=g_{1})} & =\left(-\Delta_{c2}-3\omega_{z}-i\frac{\kappa}{2}\right)\left|e_{0}\right\rangle \left\langle e_{0}\right|+\left(-\Delta_{c2}-2\omega_{z}-i\frac{\kappa}{2}\right)\left|e_{1}\right\rangle \left\langle e_{1}\right|+\mathcal{G}f_{m_{z}+1}\left(\left|e_{1}\right\rangle \left\langle e_{0}\right|+\left|e_{0}\right\rangle \left\langle e_{1}\right|\right).
\end{aligned}
\end{equation}
Each term has a clear physical meaning, for example, $\hat{H}_{{\rm NH}}^{(f=1,l=g_{0})}$ corresponds to couple from $\ket{g_0}$ to the intermediate states via the perturbation field at frequency $\omega_1$.
Notice that $\Delta_{c2}-\Delta_{c1}=\omega_z$ thus $f=2,l=e_{0}$ and $f=1,l=e_{1}$ share the same non-Hermitian Hamiltonian for the intermediate states.

Above non-Hermitian Hamiltonians take the form of $\left(\begin{array}{cc}
\tilde{E}_{0} & \mathcal{G}f_{m_{z}+1}\\
\mathcal{G}f_{m_{z}+1} & \tilde{E}_{1}
\end{array}\right)$ in the $\ket{e_{0,1}}$ basis, as a result, the inversions can be evaluated as $\frac{1}{\tilde{E}_{0}\tilde{E}_{1}-(\mathcal{G}f_{m_{z}+1})^{2}}\left(\begin{array}{cc}
\tilde{E}_{1} & -\mathcal{G}f_{m_{z}+1}\\
-\mathcal{G}f_{m_{z}+1} & \tilde{E}_{0}
\end{array}\right)$.
Then we calculate the contribution to the effective Hamiltonian from individual perturbation fields and energy levels as $\hat{H}_{{\rm eff}}^{(f,l)}=-\frac{1}{2}\hat{V}_{-} \left(\hat{H}_{{\rm NH}}^{(f,l)}\right)^{-1}\hat{V}_{+}^{(f,l)}+\mathrm{h.c.}$:
\begin{equation}
\begin{aligned}
\hat{H}_{{\rm eff}}^{(f=1,l=g_{0})} & =
\frac{\mathcal{G}^{2}\left|\alpha_{1}\right|^{2}\left(\Delta_{c1}-2\omega_{z}\right)}{\left(\Delta_{c1}-2\omega_{z}\right)^{2}+\left(\kappa/2\right)^{2}}f_{m_{z}}^{2}\left|g_{0}\right\rangle \left\langle g_{0}\right|\\
\hat{H}_{{\rm eff}}^{(f=2,l=g_{0})} & =\frac{\mathcal{G}^{3}\alpha_{1}^{*}\alpha_{2}f_{m_{z}}f_{m_{z}+1}f_{m_{z}+2}}{2\left(\Delta_{c1}+i\kappa/2\right)\left(\Delta_{c2}+i\kappa/2\right)}\left|g_{1}\right\rangle \left\langle g_{0}\right|e^{-3i\omega_{z}t}+\mathrm{h.c.}
+\frac{\mathcal{G}^{2}\left|\alpha_{2}\right|^{2}\Delta_{c2}}{\Delta_{c2}^{2}+\left(\kappa/2\right)^{2}}f_{m_{z}}^{2}\left|g_{0}\right\rangle \left\langle g_{0}\right|\\
\hat{H}_{{\rm eff}}^{(f=1,l=g_{1})} & =\frac{\mathcal{G}^{3}\alpha_{2}\alpha_{1}^{*}f_{m_{z}}f_{m_{z}+1}f_{m_{z}+2}}{2\left(\Delta_{c1}+i\kappa/2\right)\left(\Delta_{c2}+i\kappa/2\right)}\left|g_{0}\right\rangle \left\langle g_{1}\right|e^{3i\omega_{z}t}+\mathrm{h.c.}
+\frac{\mathcal{G}^{2}\left|\alpha_{1}\right|^{2}\Delta_{c1}}{\Delta_{c1}^{2}+\left(\kappa/2\right)^{2}}f_{m_{z}+2}^{2}\left|g_{1}\right\rangle \left\langle g_{1}\right|\\
\hat{H}_{{\rm eff}}^{(f=2,l=g_{1})} & =\frac{\mathcal{G}^{2}\left|\alpha_{2}\right|^{2}\left(\Delta_{c2}+2\omega_{z}\right)}{\left(\Delta_{c2}+2\omega_{z}\right)^{2}+\left(\kappa/2\right)^{2}}f_{m_{z}+2}^{2}\left|g_{1}\right\rangle \left\langle g_{1}\right|.
\end{aligned}
\end{equation}
Notice that we ignore the coupling terms $\ket{g_1}\bra{g_0}$ in $\hat{H}_{{\rm eff}}^{(f=1,l=g_{0})}$ and $\hat{H}_{{\rm eff}}^{(f=2,l=g_{1})}$, as they oscillate with 
$e^{3i\omega_z t}$ and are therefore negligible in the atomic rotating frame descried by the Hamiltonian  $\omega_z \hat{J}_z$. Also we assume $\mathcal{G} \sqrt{N} \ll \Delta_{c1,2}$ in the experiment,  allowing us to neglect the $(\mathcal{G}f_{m_{z}+1})^{2}$ terms in the denominators (they contribute to even high-order terms like $\hat{J}^2_-\hat{J}^2_+$). 

A similar treatment can be applied to each Dicke state $\ket{m_z}$, and summing over $m_z$  yields the effective Hamiltonian:
\begin{equation}
\begin{aligned}
\hat{H}_{{\rm eff}}
&\approx\mathcal{G}^{3}\alpha_{1}^{*}\alpha_{2}\mathrm{Re}\frac{1}{\left(\Delta_{c1}+i\kappa/2\right)\left(\Delta_{c2}+i\kappa/2\right)}\hat{J}_{+}^{3}e^{-3i\omega_{z}t}+\mathrm{h.c.}\\
 & + \mathcal{G}^{2}\left(\frac{\left|\alpha_{1}\right|^{2}\Delta_{c1}}{\Delta_{c1}^{2}+\left(\kappa/2\right)^{2}}+\frac{\left|\alpha_{2}\right|^{2}\left(\Delta_{c2}+2\omega_{z}\right)}{\left(\Delta_{c2}+2\omega_{z}\right)^{2}+\left(\kappa/2\right)^{2}}\right)\hat{J}_{+}\hat{J}_{-}\\
&+\mathcal{G}^{2}\left(\frac{\left|\alpha_{1}\right|^{2}\left(\Delta_{c1}-2\omega_{z}\right)}{\left(\Delta_{c1}-2\omega_{z}\right)^{2}+\left(\kappa/2\right)^{2}}+\frac{\left|\alpha_{2}\right|^{2}\Delta_{c2}}{\Delta_{c2}^{2}+\left(\kappa/2\right)^{2}}\right)\hat{J}_{-}\hat{J}_{+} \\
&+\omega_{z}\hat{J}_{z} \label{eq:10}
\end{aligned}
\end{equation}
The first line describes the 3-body interaction term. In the frame  co-rotating with the atomic coherence, this term becomes time-independent.
The second and third lines represent the exchange interaction, which vanishes under the experimental conditions $\Delta_{c1}=-\Delta_{c2}$ and $\left|\alpha_{1}\right|=\left|\alpha_{2}\right|$. 

The effective jump operators are given by:
\begin{equation}
\begin{aligned}
\hat{L}_{\mathrm{eff}} & =\mathcal{G}\sqrt{\kappa}\alpha_1\left(\frac{\hat{J}_{+}e^{-i\omega_{1}t}}{-\Delta_{c1}+2\omega_{z}-i\frac{\kappa}{2}}+\frac{\hat{J}_{-}e^{-i\omega_{1}t}}{-\Delta_{c1}-i\frac{\kappa}{2}}\right)\\
&+ \mathcal{G}\sqrt{\kappa}\alpha_2\left(\frac{\hat{J}_{+}e^{-i\omega_{2}t}}{-\Delta_{c2}-i\frac{\kappa}{2}}+\frac{\hat{J}_{-}e^{-i\omega_{2}t}}{-\Delta_{c2}-2\omega_{z}-i\frac{\kappa}{2}}\right)\\
 & +\frac{\mathcal{G}^{2}}{\left(\Delta_{c1}+i\kappa/2\right)\left(\Delta_{c2}+i\kappa/2\right)}\sqrt{\kappa}\left(\alpha_{1}e^{-i\omega_{1}t}\hat{J}_{-}^{2}+\alpha_{2}e^{-i\omega_{2}t}\hat{J}_{+}^{2}\right).
\end{aligned}
\end{equation}
The first and second lines represent the superradiance process, where the atom absorbs a pump photon and subsequently the cavity photon leaks out of the cavity~\cite{ThompsonMomentumExchange2024}. The third line arises from a higher-order process, where the atom first absorbs a pump photon, then evolves under the operator $\hat{V}_0$, and finally, the cavity photon leaks out of the cavity. This process is typically of order $\mathcal{G}^2/\Delta_{c1,2}^2$, which is much weaker than the superradiance and can be neglected under current experimental conditions.

Since in  a frame   rotating with the atomic coherence, each term oscillates at distinct frequency, we can  treat them as independent dissipation channels without interference. Under the condition $\Delta_{c1}=-\Delta_{c2}$ and $\left|\alpha_{1}\right|=\left|\alpha_{2}\right|$, the superradiance processes become balanced, yielding $\hat{L}_+ = \sqrt{\Gamma} \hat{J}_+$ and $\hat{L}_- = \sqrt{\Gamma} \hat{J}_-$ where $\Gamma\approx \mathcal{G}^2 \kappa |\alpha_1|^2/(\Delta_{c1}^2+(\kappa/2)^2) $. We can then compute the Heisenberg equations of motion of  some spin observables:
\begin{equation}
\begin{aligned}
\frac{d}{dt} \braket{\hat{J}_x}&=\cdots - \Gamma \braket{\hat{J}_x}\\
\frac{d}{dt} \braket{\hat{J}_y}&=\cdots - \Gamma \braket{\hat{J}_y}\\
\frac{d}{dt} \braket{\hat{J}_z}&=\cdots - 2\Gamma \braket{\hat{J}_z}\\
\frac{d}{dt} \braket{\hat{J}_x^2}&=\cdots + 2\Gamma \left(\braket{\hat{J}^2_z} - \braket{\hat{J}^2_x}\right)\\
\frac{d}{dt} \braket{\hat{J}_y^2}&=\cdots + 2\Gamma \left(\braket{\hat{J}^2_z} - \braket{\hat{J}^2_y}\right)\\
\frac{d}{dt} \braket{\hat{J}_z^2}&=\cdots + 2\Gamma \left(\braket{\hat{J}^2} - 3\braket{\hat{J}^2_z}\right).
\end{aligned}
\end{equation}
Here, $\cdots$ represents the contributions from the unitary dynamics. It is worth noting that due to the balance in the superradiance processes, the collective enhancement for superradiance disappears, resulting in a much slower time scale (by $O(N)$) compared to the collective dynamics.

A similar approach can be applied to derive the 4-body interaction shown in Fig.~\ref{fig5}, where we must consider three intermediate states coupled via the operator $\hat{V}_0$, and the non-Hermitian Hamiltonian $\hat{H}_{\rm NH}$ becomes a $3 \times 3$ matrix. Higher-order processes, such as terms like $\hat{J}^2_-\hat{J}^2_+$, can occur in the system. However, only the term $\hat{J}_+^4$ accounts for the four-cycle oscillation observed in Fig.~\ref{fig4}. On the other hand, the non-symmetrical pump scheme in this case leads to unbalanced superradiance, which becomes significant, as indicated by the observed shift of the signal away from $J_z=0$.
 
\subsection{Average Hamiltonian theory \label{sec:ave}}
In this section, we present an alternative derivation based on well-known time-averaging Hamiltonian theory~\cite{james2007effective} extending to higher-order, which yields consistent results when setting $\kappa=0$.
We start with a general time-dependent Hamiltonian $\hat{H}_{I}=\sum_{i}\hat{h}_{i}^{\dagger}e^{if_{i}t}+h.c.$ ($f_i>0$) and unitary time evolution operator $\hat{U}(t,0)$ satisfying $i\hbar\frac{\partial}{\partial t}\hat{U}(t,0)=\hat{H}_{I}(t)\hat{U}(t,0)$, the time-averaged Hamiltonian is expressed as~\cite{james2007effective}:
\begin{equation}
H_{\rm eff} = \frac{1}{2} \{ \mathcal{H}_{{\rm eff}}(t) + \mathcal{H}^\dagger_{{\rm eff}}(t)\} \quad\mathcal{H}_{{\rm eff}}(t)=\{\overline{\hat{H}_{I}(t)\hat{U}(t,0)}\}\{\overline{\hat{U}(t,0)}\}^{-1}.
\end{equation}
Here $\overline{\cdots}$ denotes a time average, and the fast-rotating terms are discarded.
The time-order expansion for $\hat{U}(t,0)$ is given by:  
\begin{equation}
\begin{aligned}
\hat{U}(t,0) & =\hat{T}\exp[-i\int_{0}^{t}\hat{H}_{I}(t^{\prime})dt^{\prime}]\\
 & =I+\frac{1}{i\hbar}\int_{0}^{t}\hat{H}_{I}(t_{1})dt_{1}-\frac{1}{\hbar^{2}}\int_{0}^{t}dt_{1}\int_{0}^{t_{1}}dt_{2}\hat{H}_{I} (t_{1})\hat{H}_{I}(t_{2})+\cdots
\end{aligned}
\end{equation}
and we can define the first and second-order propergators,
\begin{equation}
\hat{U}_{1}(t) =\frac{1}{i\hbar}\int_{0}^{t}\hat{H}_{I} \quad 
 \hat{U}_{2}(t) = -\frac{1}{\hbar^{2}}\int_{0}^{t}dt_{1}\int_{0}^{t_{1}}dt_{2}\hat{H}_{I}(t_{1})\hat{H}_{I}(t_{2}) 
\end{equation}
% with the first and second-order term
% \begin{equation}
% \begin{aligned}
% \hat{U}_{1}(t) & =\frac{1}{i\hbar}\int_{0}^{t}\hat{H}_{I}(t_{1})dt_{1}=\sum_{i}\hat{h}_{i}^{\dagger}\frac{1-e^{i f_i t}}{ f_i }-\hat{h}_{i}\frac{1-e^{-i f_i t}}{ f_i }\\
% \hat{U}_{2}(t) & = -\frac{1}{\hbar^{2}}\int_{0}^{t}dt_{1}\int_{0}^{t_{1}}dt_{2}\hat{H}_{I}(t_{1})\hat{H}_{I}(t_{2}) \\
% & =\sum_{ij}\hat{h}_{i}^{\dagger}\hat{h}_{j}^{\dagger}[\frac{1}{ f_i ( f_i +\omega_{j})}+\frac{e^{i( f_i +\omega_{j})t}}{\omega_{j}( f_i +\omega_{j})}-\frac{e^{i f_i t}}{ f_i \omega_{j}}]+\hat{h}_{i}\hat{h}_{j}[\frac{1}{ f_i ( f_i +\omega_{j})}+\frac{e^{-i( f_i +\omega_{j})t}}{\omega_{j}( f_i +\omega_{j})}-\frac{e^{-i f_i t}}{ f_i \omega_{j}}]\\
%  & +\sum_{i\neq j}\hat{h}_{i}^{\dagger}\hat{h}_{j}[\frac{1}{ f_i ( f_i -\omega_{j})}-\frac{e^{i( f_i -\omega_{j})t}}{\omega_{j}( f_i -\omega_{j})}+\frac{e^{i f_i t}}{ f_i \omega_{j}}]+\hat{h}_{i}\hat{h}_{j}^{\dagger}[-\frac{1}{ f_i (- f_i +\omega_{j})}+\frac{e^{i(- f_i +\omega_{j})t}}{\omega_{j}(- f_i +\omega_{j})}+\frac{e^{-i f_i t}}{ f_i \omega_{j}}]\\
%  & -\sum_{i}\hat{h}_{i}^{\dagger}\hat{h}_{i}\frac{1+it f_i -e^{it f_i }}{ f_i ^{2}}-\hat{h}_{i}\hat{h}_{i}^{\dagger}\frac{1-it f_i -e^{-it f_i }}{ f_i ^{2}} .
% \end{aligned}
% \end{equation}
Moreover, the properties $\hat{H}_{I}^{\dagger}(t)=\hat{H}_{I}(t)$ and $\hat{U}_{1}^{\dagger}(t)=-\hat{U}_{1}(t)$ hold, and under the time average $\overline{\hat{H}_{I}(t)}=0$, $\overline{\hat{U}_{1}(t)}=\sum_{i}\frac{\hat{h}_{i}^{\dagger}-\hat{h}_{i}}{ f_i }$.

Retaining terms up to third order in $H_I(t)$, we find
\begin{equation}
\begin{aligned}
\mathcal{H}_{{\rm eff}} & \approx\overline{\hat{H}_{I}(t)\{I+\hat{U}_{1}(t)+\hat{U}_{2}(t)\}}\times\overline{\{I+\hat{U}_{1}(t)+\hat{U}_{2}(t)\}}{}^{-1}\\
 % & \approx\overline{\hat{H}_{I}(t)\{I+\hat{U}_{1}(t)+\hat{U}_{2}(t)\}}\times\overline{\{I-\hat{U}_{1}(t)-\hat{U}_{2}(t)+\hat{U}_{1}^{2}(t)\}}\\
 % & \approx\overline{\hat{H}_{I}(t)}+\overline{\hat{H}_{I}(t)\hat{U}_{1}(t)}-\overline{\hat{H}_{I}(t)}\times\overline{\hat{U}_{1}(t)}+\overline{\hat{H}_{I}(t)\hat{U}_{2}(t)}-\overline{\hat{H}_{I}(t)}\times\overline{\hat{U}_{2}(t)}+\overline{\hat{H}_{I}(t)}\times\overline{\hat{U}_{1}^{2}(t)}-\overline{\hat{H}_{I}(t)\hat{U}_{1}(t)}\times\overline{\hat{U}_{1}(t)}\\
 & \approx\overline{\hat{H}_{I}(t)\hat{U}_{1}(t)}+\overline{\hat{H}_{I}(t)\hat{U}_{2}(t)}-\overline{\hat{H}_{I}(t)\hat{U}_{1}(t)}\times\overline{\hat{U}_{1}(t)}.
\end{aligned}
\end{equation}
Thus, the effective Hamiltonian is expressed as
\begin{equation}
\begin{aligned}
H_{{\rm eff}}=\frac{1}{2}(\mathcal{H}_{{\rm eff}}+\mathcal{H}_{{\rm eff}}^{\dagger})
=\frac{1}{2}\{\overline{[\hat{H}_{I}(t),\hat{U}_{1}(t)]}+\overline{\hat{H}_{I}(t)\hat{U}_{2}(t)}+\overline{\hat{U}_{2}^{\dagger}(t)\hat{H}_{I}(t)}-\overline{\hat{H}_{I}(t)\hat{U}_{1}(t)}\times\overline{\hat{U}_{1}(t)}-\overline{\hat{U}_{1}(t)}\times\overline{\hat{U}_{1}(t)\hat{H}_{I}(t)}\}. \label{eq:Htav}
\end{aligned}
\end{equation}
The first term corresponds to the second-order contribution ($\mathcal{G}^2$) while the remaining terms represent the third-order contribution ($\mathcal{G}^3$) in the effective Hamiltonian.

Back to our case, in the interaction picture of $\hat{H}_0= \omega_c \hat{b}^\dagger \hat{b} + \omega_{z}\hat{J}_{z}$,  the atom-cavity coupling Hamiltonian is expressed as:
\begin{equation}
\begin{aligned}
\hat{H}_{I}/\hbar & =\mathcal{G} \hat{a}^{\dagger}\hat{a}(\hat{J}_{+}e^{i\omega_{z}t}+\hat{J}_{-}e^{-i\omega_{z}t})\\
 & =\mathcal{G}(\hat{b}^{\dagger}e^{i\omega_c t}+\alpha_{1}^{*}e^{i\omega_{1}t}+\alpha_{2}^{*}e^{i\omega_{2}t})(\hat{b}e^{-i\omega_c t}+\alpha_{1}e^{-i\omega_{1}t}+\alpha_{2}e^{-i\omega_{2}})(\hat{J}_{+}e^{i\omega_{z}t}+\hat{J}_{-}e^{-i\omega_{z}t}). \label{eq:hI}
\end{aligned}
\end{equation}
Then we can decomposed $\hat{H}_I(t)$ via:
\begin{equation}
\begin{aligned}
\hat{h}_{0}^{\dagger} & =\mathcal{G}(\hat{b}^{\dagger}\hat{b}+|\alpha_{1}|^{2}+|\alpha_{2}|^{2})\hat{J}_{+}\quad f_{0}=\omega_{z}\\
\hat{h}_{1}^{\dagger} & =\mathcal{G}\alpha_{1}\hat{b}^{\dagger}\hat{J}_{-}\quad f_1=-\Delta_{c1}\equiv\omega_{c}-\omega_{1}-\omega_{z}\\
\hat{h}_{2}^{\dagger} & =\mathcal{G}\alpha_{2}^{*}\hat{b}\hat{J}_{-}\quad f_2=\Delta_{c2}\equiv\omega_{2}-\omega_{c}-\omega_{z} \\
\hat{h}_{3}^{\dagger} & = \mathcal{G}\alpha_{2}^{*}\alpha_{1}\hat{J}_{+}\quad f_3 = 4\omega_z \\
\hat{h}_{4}^{\dagger} & = \mathcal{G}\alpha_{2}^{*}\alpha_{1}\hat{J}_{-}\quad f_4 = 2\omega_z \\
\hat{h}_{5}^{\dagger} & =\mathcal{G}\alpha_{1}\hat{b}^{\dagger}\hat{J}_{+}\quad f_5=-\Delta_{c1}+2\omega_z = \omega_{c}-\omega_{1}+\omega_{z} \\
\hat{h}_{6}^{\dagger} & =\mathcal{G}\alpha_{2}^{*}\hat{b}\hat{J}_{+}\quad f_6
=\Delta_{c2}+2\omega_z =\omega_{2}-\omega_{c}+\omega_{z}.
\end{aligned}
\end{equation}
The physical interpretation of each term is straightforward; for instance, $\hat{h}_{1,5}$ represents the blue and red modulation sideband for the first pump.  In our pump scheme, we set $\omega_{2}-\omega_{1}=3\omega_{z}$, thus
$f_{2}+f_{1}=\omega_{z}$ and $\left|f_2-f_1\right|\equiv\delta\ll\omega_{z},\Delta_{c1,2}$.
All other frequency components vanish under time averaging, except for $f_2 - f_1$.

Next, we apply the formula introduced in Eq.~\eqref{eq:Htav}. For the second-order terms, we obtain
\begin{equation}
\begin{aligned}
\frac{1}{2}\overline{[\hat{H}_{I}(t),\hat{U}_{1}(t)]} & =\frac{1}{2}[\overline{\sum_{i}\hat{h}_{i}^{\dagger}e^{i f_i t}+\hat{h}_{i}e^{-i f_i t},-\sum_{i}\hat{h}_{i}^{\dagger}\frac{e^{i f_i t}}{ f_i }+\hat{h}_{i}\frac{e^{-i f_i t}}{ f_i }}]\\
 & =\sum_{i}\frac{[\hat{h}_{i}^{\dagger},\hat{h}_{i}]}{ f_i }+\frac{[\hat{h}_{1}^{\dagger},\hat{h}_{2}]e^{i(f_{1}-f_{2})t}}{\bar{f}_{12}}+\frac{[\hat{h}_{2}^{\dagger},\hat{h}_{1}]e^{i(f_{2}-f_{1})t}}{\bar{f}_{12}}\\
 &\approx \frac{\mathcal{G}^{2}|\alpha_{2}|^{2}}{\Delta_{c2}}\hat{J}_{-}\hat{J}_{+} + \frac{\mathcal{G}^{2}|\alpha_{2}|^{2}}{\Delta_{c2}+2\omega_z}\hat{J}_{+}\hat{J}_{-} + \frac{\mathcal{G}^{2}|\alpha_{1}|^{2}}{\Delta_{c1}}\hat{J}_{+}\hat{J}_{-} + \frac{\mathcal{G}^{2}|\alpha_{1}|^{2}}{\Delta_{c1}-2\omega_z}\hat{J}_{-}\hat{J}_{+},
\end{aligned}
\end{equation}
where $\frac{2}{\bar{f_{ij}}}=\frac{1}{f_i}+\frac{1}{f_j}$ for $i\neq j$. These terms are dominated by the exchange interaction, as anticipated~\cite{ThompsonMomentumExchange2024,wilson2024entangled}. Such exchange interactions can be precisely canceled by setting 
$\Delta_{c1}=-\Delta_{c2}$ and $|\alpha_1|=|\alpha_2|$.

For the third-order term after some math, we find
\begin{equation}
\frac{1}{2}\{\overline{\hat{H}_{I}(t)\hat{U}_{1}(t)}\times\overline{\hat{U}_{1}(t)}+\overline{\hat{U}_{1}(t)}\times\overline{\hat{U}_{1}(t)\hat{H}_{I}(t)}\}
=\frac{1}{2}[\sum_{i}\frac{[\hat{h}_{i}^{\dagger},\hat{h}_{i}]}{ f_i },\sum_{j}\frac{\hat{h}_{j}^{\dagger}-\hat{h}_{j}}{\omega_{j}}],
\end{equation}
and
\begin{equation}
\begin{aligned}
\frac{1}{2}\{\overline{\hat{H}_{I}(t)\hat{U}_{2}(t)}+\overline{\hat{U}_{2}^{\dagger}(t)\hat{H}_{I}(t)}\} 
&=\frac{1}{2}  [\sum_i \frac{[\hat{h}_i^\dagger, \hat{h}_i]}{f_i}, \sum_j \frac{\hat{h}_j^\dagger - \hat{h}_j}{\omega_j}] \\
&+  \frac{f_1[[\hat{h}_0,\hat{h}_1^\dagger],\hat{h}_2^\dagger]
+ f_1[[\hat{h}_0^\dagger,\hat{h}_1],\hat{h}_2] + f_2[[\hat{h}_0,\hat{h}_2^\dagger],\hat{h}_1^\dagger]
+ f_2[[\hat{h}_0^\dagger,\hat{h}_2],\hat{h}_1] 
}{f_1f_2 (f_1+f_2)}\\
&= \frac{1}{2}  [\sum_i \frac{[\hat{h}_i^\dagger, \hat{h}_i]}{f_i}, \sum_j \frac{\hat{h}_j^\dagger - \hat{h}_j}{\omega_j}] 
+ \frac{U^3}{\Delta_{c1}\Delta_{c2}} (\alpha^*_2 \alpha_1 \hat{J}_+^3 + \alpha^*_2 \alpha_1 \hat{J}_-^3)
\end{aligned}
\end{equation}
As a result, the effective Hamiltonian is given by:
\begin{equation}
H_\mathrm{eff}= \frac{\mathcal{G}^3}{\Delta_{c1}\Delta_{c2}} (\alpha_2 \alpha_1^* \hat{J}_+^3 + \alpha_1 \alpha^*_2 \hat{J}_-^3).
\end{equation}
\section{Generation of Bragg and dressing laser tones}
We use a single laser for driving Bragg rotation and inducing the interaction to meet the phase-matching requirement~\cite{ThompsonMomentumExchange2024}. Driving the Bragg rotations between $\ket{p_0-\hbar k}$ and $\ket{p_0+\hbar k}$ requires two different optical tones on the laser separated by $\omega_z$. We realize this by first red shift the laser with one AOM and then blue shift it back with another AOM driven with two radio frequency (RF) tones at $\omega_{RF1}=2\pi\times75$~MHz and $\omega_{RF2}=2\pi\times75~\mathrm{MHz}-\omega_z$. Because the atoms accelerate under gravity, we linearly chirp the frequency separation $\omega_z/2\pi$ between the two sidebands at a rate $\delta_r=2 \pi \times 25.11$~kHz/ms to compensate the changing Doppler shift. We frequency stabilize both Bragg laser tones roughly 3~MHz  from the cavity resonance, such that both tones non-resonantly enter the cavity and drive the Bragg rotations. Since the detuning $3~\mathrm{MHz} \gg \kappa/2\pi$, we suppress any frequency noise to amplitude or phase noise conversion that would degrade the fidelity of the Bragg rotations. For all of the Bragg pulses applied in the interferometer sequence, the Rabi frequency is $8~$kHz, giving a $\pi$-pulse duration of $63.~\mu$s.

% XXI am not following the notation here.  Can you please just write down the 3-photon first and then the 4-photon? XX

For driving the 3-body interactions, we need two dressing laser tones separated by $3\omega_z$ while being phase-coherent with respect to the Bragg coupling. To achieve that, we first triple the frequency of $\omega_{RF2}$ and double the frequency of $\omega_{RF1}$.  We then mix the multiplied signals together to generate a third tone at frequency $\omega_{RF3}=3\omega_{RF2}-2\omega_{RF1}=2\pi\times 75~\mathrm{MHz}-3\omega_z$. We combine this new RF tone with $\omega_{RF1}=2\pi\times 75$~MHz to drive an AOM (acousto-optic modulator) to create two optical laser tones separated by $3\omega_z$. Since this new RF frequency tone is generated from the same frequency sources used to drive the Bragg transitions $\omega_{RF1}$ and $\omega_{RF2}$, the interaction phase is thus phase stabilized to the Bragg coupling. 

The driving laser tones for 4-body interaction can be engineered in a similar way. In this case, we quadrupole the frequency of $\omega_{RF2}$, triple the frequency of $\omega_{RF1}$ and mix them together to get $\omega_{RF3}=4\omega_{RF2}-3\omega_{RF1}=2\pi\times 75~\mathrm{MHz}-4\omega_z$. We then combine this RF tone with $\omega_{RF1}$ and drive an AOM to create two optical laser tones separated by $4\omega_z$ for driving the 4-body interactions with a well defined phase relationship to the Bragg coupling.

In every run of the experiment, the second RF tones $\omega_{RF2}$ starts chirping with respect to the first RF tone $\omega_{RF2}$ at $t=0$ when the atoms start falling. For the first Bragg pulse that creates the superposition of momentum states, the relative phase between the two RF tones defines the phase for the Bragg coupling $\phi_B = \phi_{RF2}-\phi_{RF1}= \omega_z t + \pi \delta_r t^2$. For driving the interaction, the two dressing laser tones are created by the RF tones $\omega_{RF1}$ and $\omega_{RF3}$ applied at the AOM. For the 3-body interaction, the relative phase between the tones is then determined by $\arg\left(\alpha_1^*\alpha_2\right) = \phi_{RF3} - \phi_{RF1} = 3\times \phi_{RF2}-\phi_{RF1}$. An additional phase shift of $\phi_d$ on $\omega_{RF2}$ will then control the relative phase of the 3-body interaction with respect to the Bragg coupling $\arg\left(\alpha_1^*\alpha_2\right) - 3 \phi_B = 3 \phi_d $.  The same logic applies to the 4-body interaction with the multiplicative fact 3 replaced by 4 in the above expressions.

\section{Momentum state selective readout}
To read out the populations in individual momentum states at the end of the experimental sequences, we begin by using microwave $\pi$-pulses to map the population back from $\ket{F=2, m_F=2}$ to $\ket{F=1, m_F=0}$.  %The process is the reverse of the above, but with an added $\pi$-pulse at the end to transfer atoms from $\ket{F=2, m_F=0}$ to $\ket{F=1, m_F=0}$.
We then apply a Raman $\pi$-pulse (two-photon Rabi frequency $4.2~$kHz) for transferring atoms from $\ket{F=1,m_F=0,p_0-\hbar k}$ to $\ket{F=2,m_F=0,p+\hbar k}$. A QND measurement is then used to measure the number of atoms or population as described above. We then blow away the $\ket{F=2}$ atoms with a laser beam resonant with the atomic $\ket{F=2\rightarrow 3'}$ transition.  We then measure the population in $p_0+\hbar k$ by applying a Raman $\pi$-pulse again with the appropriate two-photon detuning and perform a second QND measurement.  The momentum exchange interaction's residual superradiance may transfer atoms into adjacent momentum states $p_0\pm 3 \hbar k$. We iterate the above procedure to measure these populations as well. Thus, on every run of the experiment, we measure the populations in the four momentum states $p_0\pm \hbar k$ and $p_0\pm 3\hbar k$.

% \section{Interaction power}
% 12V for LO, which is 160uW. 
% -34dBm for AC signal for both pump tones
% Measure atomic probe power into the cavity, 0.6uW.
% Measure the pump power going into the EOM, then calculate the modulation. 0.35CPV when driving 3-oscillation. We are sending about 200pW of optical power into each sideband.

\section{Author contributions}

C.L.~, C.M.~and E.A.B.~contributed to the building of the experiment, conducted the experiments, and performed data analysis. C.L.~and J.K.T.~conceived and supervised the experiments. H.Z.~and A.C.~contributed to the theoretical derivation and numerical simulations supervised by A.M.R..  C.L., H.Z., A.M.R.~and J.K.T.~wrote the manuscript with all authors providing critical input and help with figures. All authors discussed the experimental implementation and results as well as contributed to the manuscript.

\section{Data availability}

All data obtained in the study are available from the corresponding author upon reasonable request.

\section{Competing interests}

The authors declare no competing interests.

% \clearpage

%

\end{document}